\documentclass[a4paper,twoside]{article}
\newcommand{\affil}[1]{$^{\rm #1}$}
\usepackage[authoryear]{natbib}
\bibpunct{(}{)}{;}{a}{}{,}
\usepackage{graphicx}
\date{} 
\title{\large\bf\flushleft A FUSE View of the Stellar Winds of Planetary Nebula Central Stars}
\author{\parbox{\textwidth}{\flushleft
\vspace{-0.5cm}
{\it 
Mart\'\i n A.\ Guerrero\affil{A}, 
Gerardo Ramos-Larios\affil{A,B}, 
and 
Derck Massa\affil{C} } \\
\vspace{0.4cm}
{\small \affil{A}\,Instituto de Astrof\'{\i}sica de Andaluc\'{\i}a, CSIC. 
Camino Bajo de Hu\'etor 50, E-18008 Granada, Spain} \\
{\small \affil{B}\,Instituto de Astronom\'\i a y Meteorolog\'\i a, ~Av.\ 
Vallarta No.\ 2602. 
~Col.\ Arcos Vallarta, CP 44130 Guadalajara, Jalisco, Mexico} \\
{\small \affil{C}\,SGT, Inc., NASA Goddard Space Flight Center, USA} \\
{\small ~~Email: mar@iaa.es}}}
\begin{document}
\twocolumn[
\begin{changemargin}{.8cm}{.5cm}
\begin{minipage}{.9\textwidth}
\vspace{-1cm}
\maketitle
%
%
\small{\bf Abstract:}
Since the \emph{IUE} satellite produced a vast collection of high-resolution 
UV spectra of central stars of planetary nebulae (CSPNe), there has not been 
any further systematic study of the stellar winds of these stars.  
The high spectral resolution, sensitivity and large number of archival 
observations in the \emph{FUSE} archive allow the study of the stellar 
winds of CSPNe in the far UV domain where lines of species spanning a 
wide excitation range can be observed.  
We present here a preliminary analysis of the P Cygni profiles 
of a sample of $\sim$60 CSPNe observed by \emph{FUSE}.  
P Cygni profiles evidencing fast stellar winds with velocities between 
200 and 4,300~km~s$^{-1}$ have been found in 40 CSPNe.  
In many cases, this is the first time that fast stellar winds have been 
reported for these PNe.  
A detailed study of these far-UV spectra is on-going.

\medskip{\bf Keywords:} 

\medskip
\medskip
\end{minipage}
\end{changemargin}
]
\small

\section{Introduction}

Fast stellar winds driven by radiation pressure are characteristics of 
the central stars of planetary nebulae (CSPNe).
These stellar winds, with terminal velocities ($v_\infty$) up to 
4,000~km~s$^{-1}$, carry large amounts of energy and momentum and 
interact with the slow, 5-30~km~s$^{-1}$ \citep{ELT88}, dense wind 
of the Asymptotic Giant Branch (AGB) phase.  
This interaction plays an important role in the shaping and evolution of PNe, 
as recognized by the canonical Interacting Stellar Wind model of formation of 
PNe \citep{KPF78,B87}.

The fast stellar winds in CSPNe can be discovered through the P Cygni 
profiles of lines in the UV of high excitation ions. 
The International Ultraviolet Explorer \emph{IUE} satellite obtained 
useful UV spectra in the 1,150-3,350 \AA\ range for $\sim$160 CSPNe 
(Patriarchi \& Perinotto 1991, and references therein).  
A significant fraction of these CSPNe presented P Cygni profiles in the 
N~{\sc v} $\lambda\lambda$1239,1243 \AA, C~{\sc iv} $\lambda\lambda$1548,1551 
\AA, and O~{\sc v} $\lambda$1371 \AA\ lines, among others.  
These P Cygni profiles implied fast stellar winds with edge 
velocities ranging from 600 to 3,500~km~s$^{-1}$ \citep{CSP85}.

Launched in June 1999, the Far Ultraviolet Spectroscopic Explorer 
(\emph{FUSE}) opened a new window in the far-UV range of the spectrum 
from 905 \AA\ to 1,195 \AA.  
This spectral range includes information on a variety of resonance 
lines of high excitation species (O~{\sc vi}, P~{\sc v}, Si~{\sc iv}, 
C~{\sc iii}, ...) that can be present in the spectra of CSPNe.  
The occurrence of P Cygni profiles of these lines and their properties 
($v_\infty$, variability, main ionization stage, ...) is a valuable 
tool to assess the importance of stellar winds in the formation of 
PNe.  
We have therefore started a program aimed to use the high-resolution 
spectra of CSPNe in the final archive of the \emph{FUSE} mission to 
investigate stellar winds in CSPNe.  
Here, we present preliminary results of this on-going project.

\section{Results}

The final \emph{FUSE} archive includes high-resolution spectra for 
$\sim$90 CSPNe.  
The inspection of these spectra has revealed P Cygni profiles indicative 
of stellar winds in 40 PNe.
For a dozen of them, this is the first time that fast stellar winds have 
been reported.  
The CSPNe with useful \emph{FUSE} observations that do not show 
evidence of P Cygni profiles overimposed on their stellar continuum 
are: A\,7,  A\,31,  A\,35,  A\,39,  DeHt\,2, HDW\,4,  Hen\,2-86, 
Hen\,2-138, Hen\,3-1357, K\,1-26, K\,2-2, NGC\,1360, NGC\,3132, 
NGC\,3587, NGC\,7293, Ps\,1, PuWe\,1, and Sh\,2-174.  
These are either (a) CSPNe of high $T_{\rm eff}$ and $g$ at the center 
of old PNe (e.g., NGC\,7293), i.e., these CSPNe are subdwarfs that have 
already initiated their evolution towards the white dwarf phase, or (b) 
post-AGB stars at the center of young PNe (e.g., Hen\,3-1357) that have 
not developed yet a stable wind or whose $T_{\rm eff}$ is not high enough 
to excite these emission lines in the stellar wind.

We list in Table~1 the CSPNe exhibiting P Cygni profiles and their edge 
velocities.  
Previous information obtained by \emph{IUE} has been incorporated into this 
table to allow the straightforward comparison with the new \emph{FUSE} 
measurements.   
The comparison between \emph{IUE} and \emph{FUSE} data shows general 
agreement, but there are a few CSPNe where this is not the case.  
The poorer spectral resolution of the \emph{IUE} data (e.g., NGC\,2392) 
or the difficulties in the determination of the edge velocity in CSPNe 
severely affected by H$_2$ and atomic absorptions (e.g., NGC\,6826) may 
be blamed for these differences.  
There are, however, CSPNe for which the different edge velocities 
between \emph{IUE} and \emph{FUSE} data seem real (e.g., IC\,418).

\begin{figure}[ht!]
\begin{center}
\includegraphics[bb=25 285 655 565, scale=0.39, angle=0]{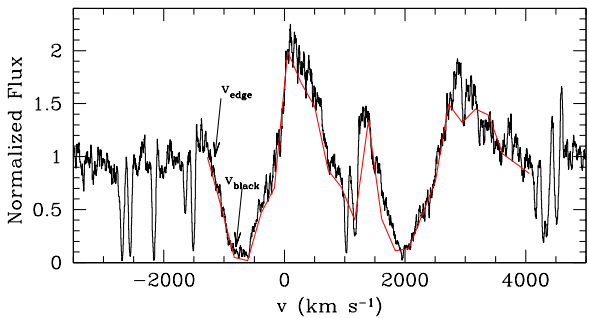}
\caption{
PB\,8 P Cygni profile of the P~{\sc v} $\lambda\lambda$1118,1128\AA\ 
line.  
The edge and black velocities are marked. 
The red curve corresponds to a SEI fit of the line profiles. 
}
\label{fig1}
\end{center}
\end{figure}

We note that the terminal velocity of a stellar wind is usually determined 
from the blue edge velocity (i.e., the maximum velocity at which the P Cygni 
profile joins back to the stellar continuum).  
Consequently, we have provided in this work the edge velocity to allow 
a fair comparison with works in the literature that used \emph{IUE} 
data (e.g., Patriarchi \& Perinotto 1991).  
Some authors, however, argue that the so-called black velocity describes 
better $v_\infty$.  
A detailed modeling using a SEI (Sobolev with Exact Integration) code 
results in a more accurate determination of $v_\infty$. 
This method is illustrated in Figure~\ref{fig1} for the CSPN of PB\,8.  
The terminal velocity of the fit, $\sim$1,200~km~s$^{-1}$, is very 
similar in this case to the edge velocity of 1,250~km~s$^{-1}$ given 
in Table~1.

As shown in Table~1, many CSPNe have P Cygni profiles of a variety 
of resonance lines of species of different excitation levels.  
A close examination of the P Cygni profiles of the different lines 
for every single CSPN reveals cases when the line profiles are 
dramatically different.  
Moreover, as in the case of the comparison between the edge velocities 
derived from \emph{IUE} and \emph{FUSE} data, there are notable cases 
of CSPNe for which different edge velocities are associated with different 
lines in the \emph{FUSE} spectral range.

\begin{figure}[ht!]
\begin{center}
\includegraphics[bb=25 285 655 565, scale=0.39, angle=0]{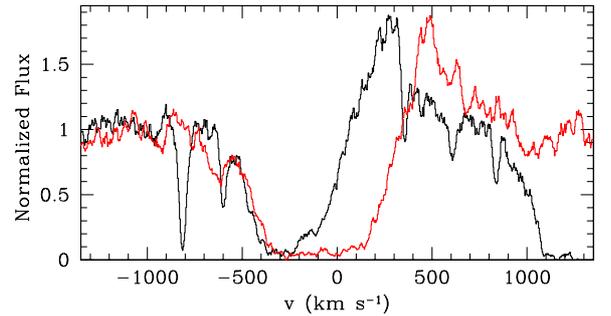}
\caption{
Cn\,3-1 P Cygni profiles of the C~{\sc iii} $\lambda$1175\AA\ (red) 
and S~{\sc iv} $\lambda\lambda$1073.0,1073.5\AA\ (black) lines. 
Note the interstellar/circumstellar absorptions in the blue edge of the 
S~{\sc iv} P Cygni profile. 
}
\label{fig2}
\end{center}
\end{figure}

First, we shall note that the shape of the P Cygni profile depends both 
on the different components and levels of the line, as well as on the 
dominant physical processes involved in its formation.  
The shape of different lines can vary owing to these factors, 
but their terminal velocities can be the same.  
This is the case for Cn~3-1 (Figure~\ref{fig2}), for which 
the profiles of the C~{\sc iii} $\lambda$1175\AA\ and 
S~{\sc iv} $\lambda\lambda$1073.0,1073.5\AA\ lines are very 
different, but the black and edge velocities are similar.  
The different profile shapes can be explained as a result of the different 
components that form these two lines: the C~{\sc iii} $\lambda$1175\AA\ 
line is a triplet which has 5 separate, closely spaced, levels, while 
the S~{\sc iv} $\lambda\lambda$1073.0,1073.5\AA\ line is one resonance 
doublet.

\begin{figure}[ht!]
\begin{center}
\includegraphics[bb=25 285 655 565, scale=0.39, angle=0]{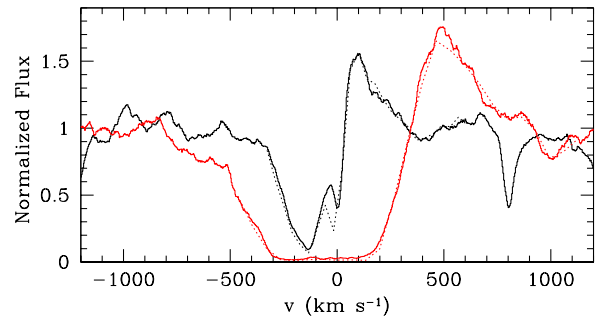}
\caption{
Hen\,2-131 P Cygni profiles of the C~{\sc iii} $\lambda$1175\AA\ (red) and 
Si~{\sc iv} $\lambda$1122\AA\ (black) lines.  
As in Figure~1, the dotted lines correspond to a SEI fits of the line 
profiles. 
}
\label{fig3}
\end{center}
\end{figure}

There are more extremes cases on which both the black and edge velocities 
and the profile shapes are notably different.  
This situation is illustrated by the P Cygni profiles of the C~{\sc iii} 
$\lambda$1175\AA\ and Si~{\sc iv} $\lambda$1122\AA\ lines of Hen\,2-131 
shown in Figure~\ref{fig3}.  
The C~{\sc iii} $\lambda$1175\AA\ line is a triplet, which can act 
much like a resonance line in dense winds, scattering radiation in 
any region wherever C$^{++}$ is present.  
On the other hand, the Si~{\sc iv} $\lambda$1122\AA\ is a line from 
a radiatively excited state.  
As the lower level of an excited state line is the upper level of a 
resonance line transition, its population depends strongly on the 
local radiation field and decreases rapidly with distance from the 
star \citep{O81}.  
Therefore, the distinct physical processes that dominate these lines 
determine not only there shapes, but also their terminal velocities.

A statistical comparison of $v_\infty$ with the stellar properties 
(spectral type, effective temperature, $T_{\rm eff}$, and gravity, $g$) 
is in progress. 
As it could be expected, stars of high gravity ($\log g > 5$) show the 
largest $v_\infty$ ($\sim$~4,000 km~s$^{-1}$).  
Two of these stars (NGC\,246 and Lo\,4) are of the PG\,1159 type. 
In contrast, stars of low gravity and effective temperature show 
low edge velocities.  
Among these CSPNe, we should mention the low edge velocities of 
several lines of Cn\,3-1, Hen\,2-131, and NGC\,2392, in the 
range 200-400 km~s$^{-1}$, whose measurement has been possible 
because the high-spectral resolution of the \emph{FUSE} data.

\section{Summary and Future Work}

Using \emph{FUSE} data, we have found evidence of fast stellar winds in 
40 CSPNe. 
For a dozen of them, this is the first time that fast stellar winds have 
been reported.  
We have determined the edge velocities of these lines, finding notable 
cases for which different edge velocities are associated to different 
lines. 
A more detail modeling using a SEI code and incorporating into the models 
the absorptions produced by circumstellar and/or interstellar H~{\sc i}, 
H$_2$, and atomic lines is on-going to determine more accurately $v_\infty$.

A statistical comparison of $v_\infty$ with stellar properties 
is also underway.  
The edge velocity is clearly correlated with the surface gravity and 
effective temperature, with the most evolved CSPNe having the fastest 
stellar winds.  
Young post-AGB stars as well as excessively evolved CSPNe do not 
show evidence of stellar winds.

\begin{table*}[h]
\centering
\caption{Edge Velocity of \emph{FUSE} and \emph{IUE} UV Lines in CSPNe }\label{tableexample}
\begin{tabular}{llr|lr}
\hline CSPN       & \emph{FUSE} Lines & Edge velocity                 & \emph{IUE} Lines               & Edge velocity    \\
                  &                                & [km~s$^{-1}$]~~~ &                                & [km~s$^{-1}$]~~~ \\
\hline
A\,30             & O~{\sc vi}, C~{\sc iii}              & 4,200~~~~~ & N~{\sc v}, O~{\sc v}, C~{\sc iv}     & 3,400~~~~~ \\
A\,43             & O~{\sc vi}, C~{\sc iii}              & 3,900~~~~~ & $\dots$                              &            \\
A\,78             & O~{\sc vi}, C~{\sc iii}              & 4,000~~~~~ & N~{\sc v}, O~{\sc v}, C~{\sc iv}     & 3,500~~~~~ \\
BD+30$^\circ$3639 & S~{\sc iv}, P~{\sc v}, Si~{\sc iv}, 
                    C~{\sc iii}                          &   850~~~~~ & N~{\sc v}, O~{\sc v}, Si~{\sc iv}, 
                                                                        C~{\sc iv}, N~{\sc iv}               & 1,000~~~~~ \\
Cn\,3-1           & S~{\sc iv}, C~{\sc iii}              &   530~~~~~ & $\dots$                              &            \\
                  & Si~{\sc iv}                          &   360~:~~~ & $\dots$                              &            \\
Hb\,7             & S~{\sc vi}                           & 1,000~~~~~ & $\dots$                              &            \\
                  & O~{\sc vi}                           & 1,500~~~~~ & $\dots$                              &            \\
Hen\,2-99         & P~{\sc v}, C~{\sc iii}               & 1,200~~~~~ & $\dots$                              &            \\
                  & S~{\sc iv}, Si~{\sc iv}              &   900~~~~~ & $\dots$                              &            \\
Hen\,2-131        & P~{\sc v}, C~{\sc iii}               &   500~~~~~ & N~{\sc v}, O~{\sc v}, Si~{\sc iv}, 
                                                                        C~{\sc iv}, N~{\sc iv}               &   850~~~~~ \\
                  & S~{\sc iv}, Si~{\sc iv}              &   300~~~~~ & $\dots$                              &            \\
Hen\,2-274        & S~{\sc iv}, C~{\sc iii}              &   600~~~~~ & $\dots$                              &            \\
Hen\,2-341        & S~{\sc vi}, O~{\sc vi}               & 1,950~~~~~ & $\dots$                              &            \\
IC\,418           & S~{\sc iv}, P~{\sc v}, 
                    C~{\sc iii} $\lambda$1175\AA         &   500~~~~~ & Si~{\sc iv}, C~{\sc iv}, N~{\sc iv}  & 1,050~~~~~ \\
                  & O~{\sc vi}, 
                    C~{\sc iii} $\lambda$977\AA          &   850~~~~~ & $\dots$                              &            \\
IC\,2149          & S~{\sc vi}, O~{\sc vi}, C~{\sc iii}  & 1,050~~~~~ & N~{\sc v}, Si~{\sc iv}, C~{\sc iv}   & 1,300~~~~~ \\
IC\,2448          & O~{\sc vi}                           & 2,550~~~~~ & $\dots$                              &            \\
IC\,2501          & S~{\sc vi}, O~{\sc vi}, P~{\sc v}    & 1,400~~~~~ & N~{\sc v}, C~{\sc iv}                & 1,280~~~~~ \\
IC\,2553          & O~{\sc vi}                           & 2,750~~~~~ & $\dots$                              &            \\
IC\,3568          & O~{\sc vi}                        & $>$1,600~~~~~ & N~{\sc v}, O~{\sc v}, C~{\sc iv}     & 1,850~~~~~ \\
IC\,4593          & P~{\sc v}, C~{\sc iii}               &   700~~~~~ & N~{\sc v}, O~{\sc iv}, Si~{\sc iv}, 
                                                                        C~{\sc iv}, N~{\sc iv}               & 1,100~~~~~ \\
                  & O~{\sc vi}                           & 1,400~~~~~ & $\dots$                              &            \\
IC\,4776          & S~{\sc vi}, O~{\sc vi}               & 2,050~~~~~ & $\dots$                              &            \\
IC\,5217          & O~{\sc vi}                           & 2,600~~~~~ & $\dots$                              &            \\
K\,1-16           & O~{\sc vi}, C~{\sc iii}              & 3,700~~~~~ & $\dots$                              &            \\
Lo\,4             & O~{\sc vi}, C~{\sc iii}              & 3,800~~~~~ & $\dots$                              &            \\
LSS\,1362         & O~{\sc vi}                           & 2,630~~~~~ & $\dots$                              &            \\
NGC\,40           & S~{\sc iv}, P~{\sc v}, C~{\sc iii}   & 1,350~~~~~ & N~{\sc v}, O~{\sc iv}, O~{\sc v}, 
                                                                        Si~{\sc iv}, C~{\sc iv}              & 1,600~~~~~ \\
                  & O~{\sc vi}                           & 1,000~:~~~ & $\dots$                              &            \\
NGC\,246          & C~{\sc iii}                          & 4,300~~~~~ & C~{\sc iv}                           & $>$3,300~~~~~ \\
                  & O~{\sc vi}                           & 3,700~~~~~ & $\dots$                              &            \\
NGC\,1535         & O~{\sc vi}, S~{\sc vi}               & 2,100~~~~~ & N~{\sc v}, O~{\sc v}                 & 2,150~~~~~ \\
NGC\,2371         & O~{\sc vi}, C~{\sc iii}              & 4,000~~~~~ & C~{\sc iv}                           & $<$3,750~~~~~ \\
NGC\,2392         & O~{\sc vi}, C~{\sc iii}              &   200~~~~~ & N~{\sc v}, N~{\sc iv}                &   600~~~~~ \\
NGC\,2867         & O~{\sc vi}, C~{\sc iii}              & 2,600~~~~~ & $\dots$                              &            \\
NGC\,5882         & O~{\sc vi}, S~{\sc vi}               & 1,950~~~~~ & N~{\sc v}, O~{\sc v}, C~{\sc iv}     & 1,525~~~~~ \\
NGC\,6058         & O~{\sc vi}                           & 2,750~~~~~ & N~{\sc v}                            &$\dots$~~~~~ \\
NGC\,6543         & S~{\sc vi}, O~{\sc vi}               & 1,900~~~~~ & N~{\sc v}, O~{\sc iv}, O~{\sc v}, 
                                                                        Si~{\sc iv}, C~{\sc iv}, N~{\sc iv}  & 1,900~~~~~ \\
                  & P~{\sc v}                            & 1,650~~~~~ & $\dots$                              &            \\
NGC\,6826         & S~{\sc vi}, O~{\sc vi}, P~{\sc v}, 
                    C~{\sc iii}                          & 1,350~~~~~ & N~{\sc v}, O~{\sc iv}, O~{\sc v}, 
                                                                        Si~{\sc iv}, C~{\sc iv}, N~{\sc iv}  & 1,600~~~~~ \\
NGC\,6891         & S~{\sc vi}, O~{\sc vi}, P~{\sc v}    & 1,400~~~~~ & N~{\sc v}, O~{\sc iv}, O~{\sc v}, 
                                                                        C~{\sc iv}, N~{\sc iv}               & 1,950~~~~~ \\
NGC\,7009         & O~{\sc vi}                           & 3,000~~~~~ & N~{\sc v}, O~{\sc v}                 & 2,750~~~~~ \\
NGC\,7094         & O~{\sc vi}, C~{\sc iii}              & 3,750~~~~~ & C~{\sc iv}                           & 3,600~~~~~ \\
NGC\,7662         & O~{\sc vi}                           & 2,550~~~~~ & $\dots$                              &            \\
PB\,6             & O~{\sc vi}                           & 3,500~~~~~ & $\dots$                              &            \\
PB\,8             & S~{\sc vi}, C~{\sc iii}              & 1,000~~~~~ & N~{\sc v}, O~{\sc iv}, O~{\sc v}, 
                                                                        Si~{\sc iv}, C~{\sc iv}, N~{\sc iv}  & 1,060~~~~~ \\
                  & O~{\sc vi}, P~{\sc v}                & 1,250~~~~~ & $\dots$                              &            \\
SwSt\,1           & O~{\sc vi}                           & 1,120~~~~~ & N~{\sc v}, O~{\sc iv}, O~{\sc v}, 
                                                                        Si~{\sc iv}, C~{\sc iv}, N~{\sc iv}  & 1,580~~~~~ \\ 
                  & P~{\sc v}                            &   800~~~~~ & $\dots$                              &            \\
                  & S~{\sc iv}, Si~{\sc iv}              &   700~~~~~ & $\dots$                              &            \\
                  & C~{\sc iii} $\lambda$1175 \AA        & 1,400~~~~~ & $\dots$                              &            \\
Vy\,2-3           & S~{\sc vi}, O~{\sc vi}               & 1,800~~~~~ & $\dots$                              &            \\
\hline
\end{tabular}
\medskip\\
\end{table*}

\section*{Acknowledgments} 
The authors acknowledge support from Ministerio de Educaci\'on y 
Ciencia (MEC), and Ministerio de Ciencia e Innovaci\'on (MICINN) 
through grants AYA2005-01495 and AYA2008-01934.



\begin{thebibliography}{}
\bibitem[Balick(1987)]{B87} 
Balick, B.\ 1987, AJ, 94, 671 
\bibitem[Cerruti-Sola \& Perinotto(1985)]{CSP85}
Cerruti-Sola, M., \& Perinotto, M.\ 1985, ApJ, 291, 237
\bibitem[Eder et al.(1988)]{ELT88}
Eder, J., Lewis, B.~M., \& Terzian, Y.\ 1988, ApJS, 66, 183
\bibitem[Kwok et al.(1978)]{KPF78} 
Kwok, S., Purton, C.~R., \& Fitzgerald, P.~M.\ 1978, ApJ, 219, L125 
\bibitem[Olson(1981)]{O81} 
Olson, G.~L.\ 1981, ApJ, 245, 1054 
\bibitem[Patriarchi \& Perinotto(1991)]{PP91} 
Patriarchi, P., \& Perinotto, M.\ 1991, A\&AS, 91, 325 
\end{thebibliography}
\end{document}